\documentclass[a4paper, 12pt]{article}   
\usepackage{amsmath}
\usepackage{amsfonts, dsfont, setspace}
\usepackage{amssymb}
\usepackage{graphicx}
\usepackage{array}
\usepackage{physics}
\usepackage{cancel}
\usepackage{wrapfig}
\usepackage{secdot}
\usepackage{indentfirst}
\usepackage{mathrsfs}
\usepackage{cancel}
\usepackage{misccorr}
\usepackage{siunitx}
\usepackage{wrapfig}	
\usepackage[english]{babel}
\usepackage[warn]{mathtext}				
\usepackage[T1, T2A]{fontenc}			
\usepackage[utf8]{inputenc} 			
\usepackage{ytableau}
\usepackage{cmap, extsizes}

\usepackage{circuitikz}
\usepackage{csquotes}

\usepackage{verbatim}

\newcommand{\mathsym}[1]{{}}
\newcommand{\unicode}[1]{{}}

\usepackage{xcolor}
\usepackage{hyperref}
\usepackage{color} 
\definecolor{linkcolor}{RGB}{0, 0, 255} 
\definecolor{urlcolor}{RGB}{255, 0, 0} 
\definecolor{citecolor}{RGB}{0, 255, 0}
\hypersetup{pdfstartview=FitH, linkcolor=linkcolor, urlcolor=urlcolor, citecolor=citecolor, colorlinks=true}

\usepackage{ragged2e}
\justifying

\usepackage{setspace}
\usepackage{geometry} 
\usepackage{titlesec}
 
\titleformat{\chapter}
    {\filcenter\Large\bfseries}  
    {\thechapter}
    {1em}{}                      

\titleformat{\section}
    {\filcenter\Large\bfseries}
    {\thesection}
    {1em}{}
 
\titleformat{\subsection}
    {\large\bfseries}
    {\thesubsection}
    {1em}{}
 
\titlespacing*{\chapter}{\parindent}{*4}{*2}
\titlespacing*{\section}{\parindent}{*4}{*2}
\titlespacing*{\subsection}{\parindent}{*4}{*2}

\usepackage{tocloft}

\usepackage{graphicx} 
\usepackage[tableposition=top]{caption}
\usepackage{subcaption}
\DeclareCaptionLabelFormat{gostfigure}{Figure #2}
\DeclareCaptionLabelFormat{gosttable}{Table #2}
\DeclareCaptionLabelSeparator{gost}{~---~}
\numberwithin{equation}{section}

\usepackage{authblk}

\usepackage[titletoc, title, header]{appendix}

\title{\bf On the relevance of quantum corrections to the matter stress-energy tensor in eternally expanding universes}
\author[1,2]{E.~T.~Akhmedov\thanks{\href{mailto:akhmedov@itep.ru}{akhmedov@itep.ru}}}
\author[1,2]{A.~V.~Anokhin\thanks{\href{mailto:anohin.av@phystech.edu}{anohin.av@phystech.edu}}}
\author[1]{K.~A.~Kazarnovskii\thanks{\href{mailto:kazarnovskiy.ka@phystech.edu}{kazarnovskiy.ka@phystech.edu}}}
\affil[1]{\textcolor{black}{Institutskii per, 9, Moscow Institute of Physics and Technology, 141700, Dolgoprudny, Russia}}
\affil[2]{\textcolor{black}{B. Cheremushkinskaya, 25, Institute for Theoretical and Experimental Physics, 117218, Moscow, Russia}}
\date{\today}

\begin{document}

\maketitle

\nopagebreak

\begin{abstract}
We study a toy-model of continuous infinite expansion of space-time with the flat start. We use as the gravitational background a conformaly flat metric with an exponentially growing factor in conformal time. We aim to clarify some properties of quantum fields in such a gravitational background. In particular, we calculate one-loop corrections to the Keldysh propagator to  verify the fact of secular growth of the occupation number and anomalous quantum average in the massless scalar field theory with selfinteractions. We perform the calculation in arbitrary dimensions with the use of the Schwinger-Keldysh technique. We get a secular growth which is not of a kinetic type. We provide some results for the case of generic interaction $\frac{\lambda}{b!}\phi^b$.
\end{abstract}

\newpage

\def\contentsname{\hfill Table of contents \hfill}
\def\refname{References}

\tableofcontents

\begin{section}{Introduction}\label{sec1}
\setcounter{figure}{0}

One of the most intriguing questions in modern cosmology is a set of properties of the very early Universe. Among those properties we have in mind the geometry of the initial Cauchy surface, basis of modes of the fields, initial Fock space state of quantum fields and etc.. To address this question now, when there is no any possible experimental test of the very early expansion, it is necessary to consider as many and as generic initial conditions as possible and see their consequences during and after the initial rapid expansion of the Universe.

The common wisdom since the original works \cite{Starobinsky:1980te}, \cite{Starobinsky:1982ee}, \cite{Linde:1981mu}, \cite{Linde:1983gd}, \cite{Guth:1980zm}, \cite{Guth:1982ec}, \cite{Albrecht:1982wi} is that at the GUT scale of inflation all other fields, except the classical gravitational background, although being quantum should be taken at the tree--level. Quantum loop effects are supposed to be contributing only the the renormalization of coupling constants, masses and other parameters of the theory. In any case, it is usually assumed that loop corrections do not drastically modify the tree-level (calculated with the uses of tree-level correlation functions) expectation value of the stress-energy tensor of matter fields. 

However, there is a substantial evidence that quantum loop corrections in non-stationary situations can play a crucial role (see \cite{Akhmedov:2021rhq} for a short recent review and \cite{Tsamis:1996qm}, \cite{Tsamis:1996qq}, \cite{Tsamis:2005hd},\cite{Krotov:2010ma}, \cite{Akhmedov:2011pj},\cite{Akhmedov:2013vka}, \cite{Akhmedov:2015xwa},\cite{Akhmedov:2019cfd},\cite{Akhmedov:2021agm}, \cite{Akhmedov:2017dih},\cite{Trunin:2021lwg}, \cite{Akhmedov:2023zfy} for incomplete list of related works). Inflationary expansion is of course represented usually by the de Sitter space-time. 
Meanwhile it is obvious that for our Universe, even if it was initially represented by de Sitter space-time with high curvature, this representation was only approximate. In fact, de Sitter isometry was violated from the very beginning either by the background itself or by the initial state of the matter fields. These observations partially clarify our statement formulated at the end of the first paragraph.

To clarify these observations in the earlier paper of our group \cite{Akhmedov:2022whm} we have considered a text book example of the short expansion between flat past and future infinities \cite{Birrell:1982ix}. In that paper we have shown that even in such a simple situation quantum loop corrections can substantially modify the tree-level expectation value of the stress-energy tensor, which is calculated in text books. Namely, we have shown that loop corrections to the stress-energy tensor are growing with time and sooner or latter (depending on the theory and initial conditions) overcome the tree-level flux. However, the question remains on whether the same situation will appear for the eternal expansion, which seems to be capable to dilute essentially any initial density of particles. In this paper we show that loop corrections to the stress-energy tensor even in the case of eternal FLRW type expansion with the flat start grow with time and can cause strong backreaction on the gravitational background. That happens even for the initial Fock space ground state.

The paper is organized as follows. In the sec. \ref{sec2} we set up the problem -- define the theory and the background. In the sec. \ref{sec3} we specify the mode functions and the initial state of the theory. In the sec. \ref{sec4} we single out leading secularly growing (with time) loop corrections to the two-point propagator that is sensitive to the state of theory. In the sec. \ref{sec5} we compare tree-level contribution and loop corrections to the stress-energy tensor. We show that while tree-level expression of the tensor vanishes at the future infinity, the loop corrections grow with time. We conclude in the sec. \ref{sec6}.

\end{section}

\begin{section}{The background geometry and setup of the problem}\label{sec2}

We consider the real scalar field theory with the action:
\begin{equation}\label{action}
    S^{free}[\phi] \equiv \int_{-\infty}^{+\infty}d^{D}x \mathscr{L}[\phi, g] = \int_{-\infty}^{+\infty} d^Dx\sqrt{|g|}\Big(\frac{g^{\mu\nu}\partial_{\mu}\phi \partial_{\nu}\phi}{2} - \frac{m^2 \phi^2}{2} - \frac{\lambda}{b!} \, \phi^b\Big),
\end{equation}
with an integer $b$ in the background metric as follows:
\begin{equation}\label{metric}
    ds^2 = C(t)\left(dt^2-d\boldsymbol{\vec{x}}^2\right), \;\; \text{where} \;\; C(t) = A+Be^{\rho t},\;\{A,B\}\;\in\; \mathds{R}^{+},\;\; \rho > 0,
\end{equation}
which describes a model of eternal expansion. To simplify equations below we make a suitable transformation of coordinates ($t^{'}=\frac{t}{\sqrt{A}}+\frac{log(\frac{B}{A})}{\rho}$) to express the metric in the following form:
\begin{equation}\label{metric2}
ds^2 = C(t)\left(dt^2-d\boldsymbol{\vec{x}}^2\right), \;\; \text{where} \;\; C(t) = 1+e^{\rho t}.
\end{equation}
At past infinity $r\rightarrow -\infty$, the metric is flat. At the same time  as $t \rightarrow +\infty$:
\begin{equation}
    ds^2 \approx  e^{\rho t}(dt^2 - d\boldsymbol{\vec{x}}^2).
\end{equation}
Then, introducing the proper time $\tau = \frac{2}{\rho}e^{\frac{\rho t}{2}}$, we obtain the asymptotic metric as follows:
\begin{equation}
    ds^2 \approx d\tau^{2} - \frac{\rho^2 \tau^2}{4}d\boldsymbol{\vec{x}}^2,\text{ as } \tau \rightarrow +\infty,
\end{equation}
which is a metric of Friedman-Lemaitre-Robinson-Walker type.

We want to calculate the flux of the produced particles at the tree and loop levels in such a background. We assume that the curvature of this space-time initially is large, although decays with time. 

Mainly we perform the calculations for the $\lambda\varphi^3$ theory. That is done to simplify the equations. Meanwhile essential physics is contained in this unstable theory. We assume that there is an additional small interaction $g\varphi^4$, $g\ll\lambda$, to stabilize the theory. But we assume that quantum corrections due to quartic term start to play a role at much later times than the corrections from $\lambda\varphi^3$ term.

\end{section}

\begin{section}{Mode expansion}\label{sec3}
\setcounter{figure}{0}
To begin with, let us find the free modes of the theory (\ref{action}) in the background (\ref{metric}). Solving the free equations of motion
\begin{equation}\label{EoM}
    \partial_{\mu}(g^{\mu\nu}\partial_{\nu}\phi\sqrt{g}) + \sqrt{g}m^2\phi = 0,
\end{equation}
in the form
\begin{equation}\label{mode}
    \phi_{\boldsymbol{\vec{k}}}(t, \boldsymbol{\vec{x}}) = f_{k}(t)e^{i\boldsymbol{\vec{k}}\boldsymbol{\vec{x}}},
\end{equation}
one obtains the equation as follows:
\begin{equation}\label{EoMexpanded}
    \left(1+e^{\rho t}\right)f^{''}(t)+\rho\left(\frac{D}{2}-1\right)e^{\rho t}f^{'}(t)+\left(1+e^{\rho t}\right)k^2f(t)+m^2\left(A+Be^{-\rho t}\right)^2f(t)=0.
\end{equation}
We hide the index $k$ of $f_{k}(t)$ for simplicity.

We cannot express solutions of (\ref{EoMexpanded}) via known special functions when $m \neq 0$ and continue with the case of massless field $m = 0$.
Making the change of the variables of the form:
\begin{equation}\label{KandD}
      K^2=\frac{k^2}{\rho^2}, \;\; d \equiv \frac{D-2}{4},
\end{equation}
we obtain the temporal part of the modes:
\begin{equation}\label{fullmodes}
    \begin{aligned}
    &f_{k}(t) =C_1 e^{-ikt} {}_2F_1\left(d-iK-i\sqrt{K^2-d^2};\;d-iK+i\sqrt{K^2-d^2};\;1-2iK;\;-e^{\rho t}\right)  + \\
    &+C_2 e^{ikt} {}_2F_1\left(d+iK-i\sqrt{K^2-d^2};\;d+iK+i\sqrt{K^2-d^2};\;1+2iK;\;-e^{\rho t}\right),\ \text{when} \ K>d,
    \end{aligned}
\end{equation}
and
\begin{equation}\label{fullmodes2}
    \begin{aligned}
    &f_{k}(t) =C_1 e^{-ikt} {}_2F_1\left(d-iK-\sqrt{d^2-K^2};\;d-iK+\sqrt{d^2-K^2};\;1-2iK;\;-e^{\rho t}\right)  + \\
    &+C_2 e^{ikt} {}_2F_1\left(d+iK-\sqrt{d^2-K^2};\;d+iK+\sqrt{d^2-K^2};\;1+2iK;\;-e^{\rho t}\right),\ \text{when}\ K<d,
    \end{aligned}
\end{equation}
where ${}_2F_{1}(a;\;b;\;c;\;z)$ is the hypergeometric function.

Now we will derive asymptotics of (\ref{fullmodes}) and (\ref{fullmodes2}) to find the in-modes. In the limit $t \rightarrow -\infty$ the modes under consideration behave as:
\begin{equation}\label{phiinmin}
    \phi_{\boldsymbol{\vec{k}}}(t \rightarrow -\infty, \boldsymbol{\vec{x}}) \approx C_1e^{-ikt+i\boldsymbol{\vec{k}}\boldsymbol{\vec{x}}} + C_2e^{ikt+i\boldsymbol{\vec{k}}\boldsymbol{\vec{x}}}.
\end{equation}
For the in-modes (single waves at past infinity) we should set $C_2 = 0$. 

\end{section}

\setcounter{figure}{0}
The coefficient $C_1$ for the in-modes is fixed from the canonical commutation relations. The field operator is:
\begin{equation}\label{modes}
    \begin{aligned}
        &\phi(t, \boldsymbol{\vec{x}}) \equiv \int\frac{d^{D-1} \boldsymbol{\vec{k}}}{(2\pi)^{D-1}}\left( a^{\dagger}_{\boldsymbol{\vec{k}}}{\phi_{\boldsymbol{\vec{k}}}^{in}}^{*}(t, \boldsymbol{\vec{x}}) + a_{\boldsymbol{\vec{k}}}\phi_{\boldsymbol{\vec{k}}}^{in}(t, \boldsymbol{\vec{x}}))\right)  = \\
        & =\int\frac{d^{D-1} \boldsymbol{\vec{k}}}{(2\pi)^{D-1}}\left(a^{\dagger}_{\boldsymbol{\vec{k}}}{f_{k}^{in}}^{*}(t)e^{-i\boldsymbol{\vec{k}}\boldsymbol{\vec{x}}} + a_{\boldsymbol{\vec{k}}}f_{k}^{in}(t)e^{i\boldsymbol{\vec{k}}\boldsymbol{\vec{x}}} \right),
    \end{aligned}
\end{equation}
and its conjugate momentum, as follows from the Lagrangian (\ref{action}), is:
\begin{equation}
    \pi(t, \boldsymbol{\vec{x}}) \equiv \frac{\partial \mathscr{L}}{\partial \partial_{0}\phi(t, \boldsymbol{\vec{x}})} = \left(1+e^{\rho t}\right)^{2d} \partial_{t}\phi(t, \boldsymbol{\vec{x}}).
\end{equation}
Here:
\begin{equation}
    \begin{aligned}
        &f_{k}^{in}(t) = C_1e^{-ikt} {}_2F_1\left(d-iK-i\sqrt{K^2-d^2},\;d-iK+i\sqrt{K^2-d^2},\;1-2iK,\;-e^{\rho t}\right), \\
        &{f_{k}^{in}}^{*}(t) = C_1^{*}e^{ikt} {}_2F_1\left(d+iK-i\sqrt{K^2-d^2},\;d+iK+i\sqrt{K^2-d^2},\;1+2iK,\;-e^{\rho t}\right).
    \end{aligned}
\end{equation}
The canonical commutation relations are:
\begin{equation}\label{commutators}
    \Big[\phi(t, \boldsymbol{\vec{x}}), \pi(t, \boldsymbol{\vec{y}}) \Big] = i\delta\left(\boldsymbol{\vec{x}} - \boldsymbol{\vec{y}}\right), \;\; \left[a_{\boldsymbol{\vec{k}}}, a^{\dagger}_{\boldsymbol{\vec{p}}} \right] = (2 \pi)^{D-1}\delta\left(\boldsymbol{\vec{k}} - \boldsymbol{\vec{p}}\right).
\end{equation}
From these relations, we find the normalization condition for $C_1$:
\begin{equation}\label{hchange}
    \Big[\partial_{t}f_{k}^{in}(t){f_{k}^{in}}^{*}(t)-\partial_{t}  {f_{k}^{in}}^{*}(t)f_{k}^{in}(t) \Big] \,  \, \left(1+e^{\rho t} \right)^{2d} = -i.
\end{equation}
This equality must be true for any moment of time because of the equations of motion (\ref{EoMexpanded}). So, setting $t \rightarrow -\infty$, we obtain:
\begin{equation}\label{C}
    C_1 = \frac{1}{\sqrt{2k}}.
\end{equation}
Moreover, one can explicitly check that the normalization condition (\ref{hchange}) is satisfied when $t \rightarrow +\infty$ and $C_1$ is given by (\ref{C}). To see that one has to use asymptotics of the in-modes at $t\rightarrow +\infty$. 

The behavior of the in-modes in the limit $t \rightarrow +\infty$ can be found using the asymptotic form of the ${}_2F_1$-hypergeometric function:
\begin{equation}
    {}_2F_{1}(a, b, c, x \rightarrow -\infty) \approx (-x)^{-a}\frac{\Gamma(c)\Gamma(b - a)}{\Gamma(c - a)\Gamma(b)} + (-x)^{-b}\frac{\Gamma(c)\Gamma(a - b)}{\Gamma(c - b)\Gamma(a)}.
\end{equation}
Thus, as $t \to \pm \infty$  for the in-modes we obtain:

\begin{equation}\label{modesas1}
    \begin{aligned}
        &f^{in}_{k}(t\rightarrow -\infty) \approx \frac{1}{\sqrt{2k}}e^{-ikt}, \\
        &f^{in}_{k}(t\rightarrow +\infty) \approx \frac{1}{\sqrt{2k}}e^{-\rho d t}\left(D_{1}(k)e^{i\omega(k) t}+D_{2}(k)e^{-i\omega(k) t}\right),\; \text{when } k>\rho d, \\
        &f^{in}_{k}(t\rightarrow +\infty) \approx\frac{1}{\sqrt{2k}}e^{-\rho d t}\left(E_{1}(k)e^{\gamma(k) t}+E_{2}(k)e^{-\gamma(k) t}\right),\; \text{when } k<\rho d.
    \end{aligned}
\end{equation}
Here, we use the following notations:
\begin{equation}\label{modesfinal1}
    \begin{aligned}
        & D_{1}(k)=\frac{\Gamma(2i\sqrt{K^2-d^2})\Gamma(1-2iK)}{\Gamma(1-d-iK+i\sqrt{K^2-d^2})\Gamma(d-iK+i\sqrt{K^2-d^2})}, \\
        & D_{2}(k)=\frac{\Gamma(-2i\sqrt{K^2-d^2})\Gamma(1-2iK)}{\Gamma(1-d-iK-i\sqrt{K^2-d^2})\Gamma(d-iK-i\sqrt{K^2-d^2})}, \\
        & E_{1}(k)=\frac{\Gamma(2\sqrt{d^2-K^2})\Gamma(1-2iK)}{\Gamma(1-d-iK+\sqrt{d^2-K^2})\Gamma(d-iK+\sqrt{d^2-K^2})}, \\
        & E_{2}(k)=\frac{\Gamma(-2\sqrt{d^2-K^2})\Gamma(1-2iK)}{\Gamma(1-d-iK-\sqrt{d^2-K^2})\Gamma(d-iK-\sqrt{d^2-K^2})}, \\
        & \omega(k) = \rho\sqrt{K^2-d^2}, \;\;\gamma(k)=\rho\sqrt{d^2-K^2}.
    \end{aligned}
\end{equation}
Where $d$ and $K$ are defined in (\ref{KandD}).

\begin{section}{Quantum loop corrections in $\frac{\lambda}{3!}\phi^3$ theory}\label{sec4}
\setcounter{figure}{0}

In this section, we calculate  loop corrections to the propagators. As the situation that we consider is non-stationary, we have to use the Schwinger-Keldysh technique where each field is described by three propagators:
\begin{equation}\label{propogators}
    \begin{aligned}
        &G^{K}(t, \boldsymbol{\vec{x}}, t^{'}, \boldsymbol{\vec{x}}^{'}) \equiv \ \frac{1}{2}\Big\langle \Big\{\phi(t, \boldsymbol{\vec{x}}), \phi(t^{'}, \boldsymbol{\vec{x}}^{'})\Big\} \Big\rangle, \\
        &G^{R}(t, \boldsymbol{\vec{x}}, t^{'}, \boldsymbol{\vec{x}}^{'}) \equiv  \theta(t-t^{'})\Big\langle \Big[\phi(t, \boldsymbol{\vec{x}}), \phi(t^{'}, \boldsymbol{\vec{x}}^{'})\Big] \Big\rangle, \\
        &G^{A}(t, \boldsymbol{\vec{x}}, t^{'}, \boldsymbol{\vec{x}}^{'}) \equiv   -\theta(t^{'}-t)\Big\langle \Big[\phi(t, \boldsymbol{\vec{x}}), \phi(t^{'}, \boldsymbol{\vec{x}}^{'})\Big] \Big\rangle. \\
    \end{aligned}
\end{equation}
We consider a spatially homogeneous initial state, in which:
\begin{equation}\label{states}
    \begin{cases}
        &\langle a^{\dagger}_{\boldsymbol{\vec{k}}} a_{{\boldsymbol{\vec{p}}}} \rangle = n_{\boldsymbol{\vec{k}}}\delta\left(\boldsymbol{\vec{k}} - \boldsymbol{\vec{p}}\right), \;\; \langle a_{\boldsymbol{\vec{k}}} a_{{\boldsymbol{\vec{p}}}} \rangle = \kappa_{\boldsymbol{\vec{k}}}\delta\left(\boldsymbol{\vec{k}} + \boldsymbol{\vec{p}}\right), \;\; \langle a^{\dagger}_{\boldsymbol{\vec{k}}} a^{\dagger}_{{\boldsymbol{\vec{p}}}} \rangle = \kappa_{\boldsymbol{\vec{k}}}^{*}\delta\left(\boldsymbol{\vec{k}} + \boldsymbol{\vec{p}}\right), \\
        & n_{\boldsymbol{\vec{k}}} = n_{-\boldsymbol{\vec{k}}}, \;\; \kappa_{\boldsymbol{\vec{k}}} = \kappa_{-\boldsymbol{\vec{k}}}.
    \end{cases}
\end{equation}
Then, spatial Fourier transformations of the propagators from eq. (\ref{propogators}) are:
\begin{equation}\label{GK0stf}
    \begin{aligned}
        &G_{0}^{K}(t, t^{'}, \boldsymbol{\vec{k}}) = \kappa_{\boldsymbol{\vec{k}}}^{*} \, f_{k}^{in}(t)f_{k}^{in}(t^{'}) + \kappa_{\boldsymbol{\vec{k}}} \, {f_{k}^{in}}^{*}(t){f_{k}^{in}}^{*}(t^{'}) +\\
        &+ \left(n_{\boldsymbol{\vec{k}}} + \frac{1}{2}\right) \, \left[{f_{k}^{in}}^{*}(t)f_{k}^{in}(t^{'}) + f_{k}^{in}(t){f_{k}^{in}}^{*}(t^{'})\right],
    \end{aligned}
\end{equation}
and
\begin{equation}\label{GR0stf}
    G_{0}^{R}(t, t^{'}, \boldsymbol{\vec{k}}) = -\theta(t-t^{'}) \, \left[{f_{k}^{in}}^{*}(t)f_{k}^{in}(t^{'}) - f_{k}^{in}(t){f_{k}^{in}}^{*}(t^{'})\right],
\end{equation}
\begin{equation}\label{GA0stf}
    G_{0}^{A}(t, t^{'}, \boldsymbol{\vec{k}}) = \theta(t^{'}-t)\left[{f_{k}^{in}}^{*}(t)f_{k}^{in}(t^{'}) - f_{k}^{in}(t){f_{k}^{in}}^{*}(t^{'})\right].
\end{equation}
As can be seen from these expressions, $G^{K}(t, \boldsymbol{\vec{x}}, t^{'}, \boldsymbol{\vec{x}}^{'})$ contains information about the state of the theory. At the same time, the tree-level Retarded and Advanced propagators $G^{R,A}(t, \boldsymbol{\vec{x}}, t^{'}, \boldsymbol{\vec{x}}^{'})$ are state independent, but dependent on the spectrum of the theory. We are mostly interested in the corrections to the Keldysh propagator, $G^{K}_{0}$, because we would like to trace the destiny of the state of the theory.

The diagrams that we will calculate in this section are shown in Fig. \ref{tadpolesphi31} and \ref{sunsetsphi3}. Diagrams from the Fig. \ref{tadpolesphi31} we will call as ''diagrams of the first type'' or tadpole diagrams. Diagrams from the Fig. \ref{sunsetsphi3} we will call as ''diagrams of the second type''.  Below, we will prove that for the initial conditions that we consider below the tadpole diagrams will result only in the mass and mode functions renormalization. Meanwhile, the occupation numbers and anomalous averages are affected only by the diagrams of the second type. They are of the main interest for us because they show the change in the state of the theory. We will calculate them in the section \ref{2type}.
\begin{figure}[ht!]
\center{\includegraphics[scale=0.5]{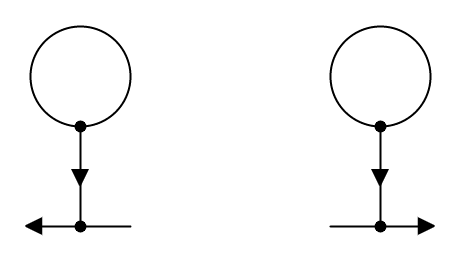}}
\caption{Diagrams of the first type.}
\label{tadpolesphi31}
\end{figure}

\subsection{Tadpole diagrams}

We start with the concise discussion of the first type or tadpole diagrams. The contribution of the tadpole diagrams from the Fig. \ref{tadpolesphi31} to the Keldysh, $G^{K}_{0}$, Retarded, $G^{R}_{0}$, and Advanced, $G^{A}_{0}$, propagators in the coordinate representation are as follows:
\begin{equation}
    \begin{aligned}
        &G^{K}_{1,\;\text{tadpoles}}(x, x^{'}) = -\frac{\lambda^2}{2}\int_{t_0}^{+\infty}dy^{0}dz^{0}\int d^{D-1}\boldsymbol{\vec{y}}d^{D-1}\boldsymbol{\vec{z}} \sqrt{g(y)g(z)}\times \\
        & \times\left[G_{0}^{R}(x, y)G_{0}^{R}(y, z)G_{0}^{K}(z, z)G_{0}^{K}(y, z^{'}) + G_{0}^{K}(x, y)G^{A}_{0}(z, y)G_{0}^{K}(z, z)G_{0}^{A}(y, x^{'})\right], \\
        &G^{R/A}_{1,\;\text{tadpoles}}(x, x^{'}) = -\frac{\lambda^2}{2}\int_{t_0}^{+\infty}dy^{0}dz^{0}\int d^{D-1}\boldsymbol{\vec{y}}d^{D-1}\boldsymbol{\vec{z}} \sqrt{g(y)g(z)}\times \\
        & \times\left[G_{0}^{R/A}(x, y)G_{0}^{R}(y, z)G_{0}^{K}(z, z)G_{0}^{R/A}(y, z^{'})\right].
    \end{aligned}
\end{equation}
These equations can be generalized to the Dyson-Schwinger equations for the resumed propagators containing only diagrams of the tadpole type:

\begin{equation}\label{tdpl}
    \begin{aligned}
        &G^{K}_{\text{tadpoles}}(x, x^{'}) = G_{0}^{K}(x, x^{'}) -\frac{\lambda^2}{2}\int_{t_0}^{+\infty}dy^{0}dz^{0}\int d^{D-1}\boldsymbol{\vec{y}}d^{D-1}\boldsymbol{\vec{z}} \sqrt{g(y)g(z)}\times \\
        & \times\Big[G_{0}^{R}(x, y)G_{0}^{R}(y, z)G_{0}^{K}(z, z)G_{\text{tadpoles}}^{K}(y, x^{'}) + \\
        &+G_{0}^{K}(x, y)G^{R}_{0}(y, z)G_{0}^{K}(z, z)G_{\text{tadpoles}}^{A}(y, x^{'})\Big], \\
        &G^{R/A}_{\;\text{tadpoles}}(x, x^{'}) = G_{0}^{R/A}(x, x^{'}) -\frac{\lambda^2}{2}\int_{t_0}^{+\infty}dy^{0}dz^{0}\int d^{D-1}\boldsymbol{\vec{y}}d^{D-1}\boldsymbol{\vec{z}} \sqrt{g(y)g(z)}\times \\
        & \times \left[G_{0}^{R/A}(x, y)G_{0}^{R}(y, z)G_{0}^{K}(z, z)G_{\text{tadpoles}}^{R/A}(y, x^{'})\right],
    \end{aligned}
\end{equation}
where $G^{K, R, A}_{1,\; \text{tadpoles}}$ are the one-loop corrections, while $G^{K, R, A}_{\text{tadpoles}}$ are the resummed expressions. If one applies the operator $\hat{L}=\partial_{\mu}g^{\mu\nu}\partial_{\nu}\sqrt{g}$  to both sides of each equation in (\ref{tdpl}), one finds the following relations:
\begin{equation}
    \begin{aligned}
        & \left\{\hat{L} + \frac{i\lambda^2}{2}\sqrt{g(x)}\int_{t_0}^{+\infty}dz^{0}\int d^{D-1}\boldsymbol{\vec{z}} \sqrt{g(z)}G_{0}^{R}(x, z)G_{0}^{K}(z, z)\right\}G_{\text{tadpoles}}^{K}(x, x^{'}) = 0, \\
        & \left\{\hat{L} + \frac{i\lambda^2}{2}\sqrt{g(x)}\int_{t_0}^{+\infty}dz^{0}\int d^{D-1}\boldsymbol{\vec{z}} \sqrt{g(z)}G_{0}^{R}(x, z)G_{0}^{K}(z, z)\right\}G_{\text{tadpoles}}^{R/A}(x, x^{'}) = \\
        &= i\delta(t-t^{'})\delta(\boldsymbol{\vec{x}}-\boldsymbol{\vec{x}}^{'}).
    \end{aligned}
\end{equation}
From these expressions one can see that besides the leading UV singularity, which are independent of the background field and the state of the theory, these expressions also contain subleading singularities which may depend on the background field and the state. Such singularities lead to a change of the modes beyond the mass renormalization  (see \cite{Akhmedov:2022whm}). Nevertheless, we will not discuss in detail such changes and assume that the physical (renormalized) mass is vanishing and all other physical quantities are taking their such values as are given in (\ref{action}). Namely, we do not discuss e.g. the interaction constant renormalization, which emerges already in the fourth order of $\lambda$ and has the form $\lambda^{(1)} \sim \lambda^3$. Such vertex renormalization diagrams appear in powers of $\lambda$ strictly greater than 2, and they can be added into diagrams of the first and second types with bold vertices. We just assume that we work with the physical (UV renormalized) mass and the interaction constant and study only infrared effects. On general physical grounds we assume that UV effects do not affect the dynamics of the background state of the theory. Meanwhile we will see below that IR effects do strongly affect the state of the theory.

\subsection{Diagramms of the second type}\label{2type}
\begin{figure}[ht!]
\center{\includegraphics[scale=0.5]{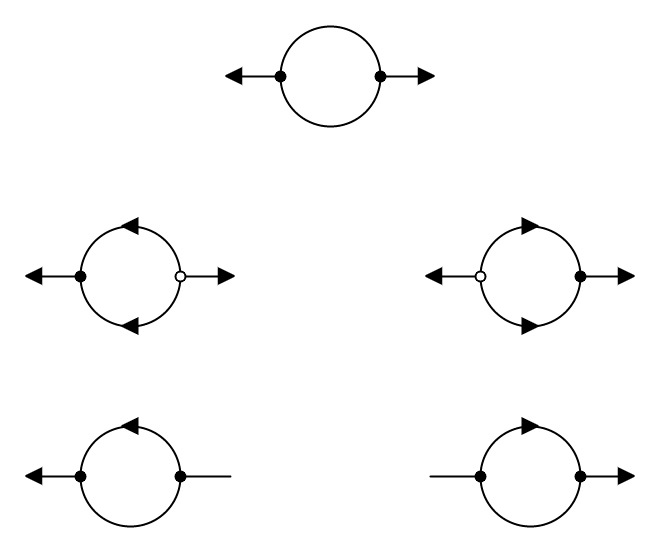}}
\caption{Diagrams of the second type.}
\label{sunsetsphi3}
\end{figure}
Now we continue with contributions of the diagrams of the second type, but only up to the second order in $\lambda$. We are not yet ready to perform the resummation of the leading contributions coming from such diagrams. The reason for that we explain below.

The $G^{K}_{0}$, $G^{R}_{0}$, and $G^{A}_{0}$ corrections in the coordinate representation, following from the Fig. \ref{sunsetsphi3} are:
\begin{equation}\label{sunsets3}
    \begin{aligned}
        & G^{K}_{1,\;\text{2-type}}(x, x^{'}) = -\frac{\lambda^2}{2}\int_{t_0}^{+\infty}dy^{0}dz^{0}\int d^{D-1}\boldsymbol{\vec{y}}d^{D-1}\boldsymbol{\vec{z}} \sqrt{g(y)g(z)}\times \\
        &\times \Big[G^{R}_{0}(x, y)\left(G^{K}_{0}(y, z)\right)^{2}G^{A}_{0}(z, x^{'})+ \\
        &+ \frac{1}{4}G^{R}_{0}(x, y)\left(G^{R}_{0}(y, z)\right)^{2}G^{A}_{0}(z, x^{'}) + \frac{1}{4}G^{R}_{0}(x, y)\left(G^{A}_{0}(y, z)\right)^{2}G^{A}_{0}(z, x^{'}) + \\
        &+ 2G^{R}_{0}(x, y)G^{K}_{0}(y, z)G^{R}_{0}(y, z)G^{K}_{0}(z, x^{'}) + 2G^{K}_{0}(x, y)G^{K}_{0}(y, z)G^{A}_{0}(y, z)G^{A}_{0}(z, x^{'})\Big], \\
        & G^{R/A}_{1,\;\text{2-type}}(x, x^{'}) = -\lambda^2\int_{t_0}^{+\infty}dy^{0} dz^{0}\sqrt{g(y^{0})g(z^{0})} \int d^{D-1}\boldsymbol{\vec{y}}d^{D-1}\boldsymbol{\vec{z}} \times \\
        &\times \left[G^{R/A}_{0}(x, y)G^{K}_{0}(y, z)G^{R/A}_{0}(y, z)G^{R/A}_{0}(z, x^{'})\right].
    \end{aligned}
\end{equation}
We are interested in the form of these corrections in the following time limit:
\begin{equation}\label{limit}
    \begin{cases}
        t + t^{'} \rightarrow +\infty \\
        t - t^{'} = \text{const},
    \end{cases}
\end{equation}
where both points of the propagators are taken to the future infinity. This limit allows one to trace the destiny of the state of the theory in the future. To simplify expressions below we introduce the notation:
\begin{equation}\label{limit2}
     t \approx t^{'} \equiv T_0 \rightarrow +\infty.
\end{equation}
It can be seen that in the limit (\ref{limit}) and (\ref{limit2}) corrections to $G^{R/A}_{1,\;\text{2-type}}$ from (\ref{sunsets3}) are negligible:
\begin{equation}\label{GR1loop}
    \begin{aligned}
        &G^{R/A}_{1,\;\text{2-type}}(x, x^{'}\;|\;t \approx t^{'}\equiv T_0) \approx -\lambda^2\int_{t_0}^{T_0}dy^{0} dz^{0}\sqrt{g(y^{0})g(z^{0})} \int d^{D-1}\boldsymbol{\vec{y}}d^{D-1}\boldsymbol{\vec{z}} \times \\
        &\times\theta(T_0-y^{0})\theta(y^{0}-z^{0})\theta(z^{0}-T_0)\times\left[G^{C}_{0}(x, y)G^{K}_{0}(y, z)G^{C}_{0}(y, z)G^{C}_{0}(z, x^{'})\right] = 0.
    \end{aligned}
\end{equation}
This happens due to the presence of the theta-functions -- the essential element of the Retarded and Advanced propagators at any loop order  \cite{Kamenev}, \cite{Berges:2004yj}. Moreover, this result can be generalized to any perturbative order due to the causality.

Thus, the most interesting contribution from our point of view is the correction to $G^{K}_{0}$ from the diagrams of the second type from eq. (\ref{sunsets3}). In terms of the spatially Fourier transformed propagators this contribution can be written as:
\begin{equation}\label{GK1loop}
    \begin{aligned}
        &G^{K}_{1,\;\text{2-type}}(t, t^{'}, \boldsymbol{\vec{k}}) = -\frac{\lambda^2}{2}\int_{t_0}^{+\infty}dy^{0} dz^{0}\sqrt{g(y^{0})g(z^{0})} \int\frac{d^{D-1}\boldsymbol{\vec{p}}}{(2\pi)^{D-1}}\frac{d^{D-1}\boldsymbol{\vec{s}}}{(2\pi)^{D-1}}\delta\left(\boldsymbol{\vec{k}}-\boldsymbol{\vec{p}}-\boldsymbol{\vec{s}}\right)\times \\
        &\times \Big[G^{R}_{0}(t, y^{0}, \boldsymbol{\vec{k}})G^{K}_{0}(y^{0}, z^{0}, \boldsymbol{\vec{p}})G^{K}_{0}(y^{0}, z^{0}, \boldsymbol{\vec{s}})G^{A}_{0}(z^{0}, t^{'}, \boldsymbol{\vec{k}}) + \\
        &+ \frac{1}{4}G^{R}_{0}(t, y^{0}, \boldsymbol{\vec{k}})G^{R}_{0}(y^{0}, z^{0}, \boldsymbol{\vec{p}})G^{R}_{0}(y^{0}, z^{0}, \boldsymbol{\vec{s}})G^{A}_{0}(z^{0}, t^{'}, \boldsymbol{\vec{k}}) + \\
        &+ \frac{1}{4}G^{R}_{0}(t, y^{0}, \boldsymbol{\vec{k}})G^{A}_{0}(y^{0}, z^{0}, \boldsymbol{\vec{p}})G^{A}_{0}(y^{0}, z^{0}, \boldsymbol{\vec{s}})G^{A}_{0}(z^{0}, t^{'}, \boldsymbol{\vec{k}}) + \\
        &+ 2G^{R}_{0}(t, y^{0}, \boldsymbol{\vec{k}})G^{K}_{0}(y^{0}, z^{0}, \boldsymbol{\vec{p}})G^{R}_{0}(y^{0}, z^{0}, \boldsymbol{\vec{s}})G^{K}_{0}(z^{0}, t^{'}, \boldsymbol{\vec{k}}) + \\
        &+ 2G^{K}_{0}(t, y^{0}, \boldsymbol{\vec{k}})G^{K}_{0}(y^{0}, z^{0}, \boldsymbol{\vec{p}})G^{A}_{0}(y^{0}, z^{0}, \boldsymbol{\vec{s}})G^{A}_{0}(z^{0}, t^{'}, \boldsymbol{\vec{k}})\Big].
    \end{aligned}
\end{equation}
From this expression, one can find corrections to the occupation number (OA) and the anomalous quantum average (AA). To do that, one should use in (\ref{GK1loop}) expressions (\ref{GK0stf}, \ref{GR0stf}, \ref{GA0stf}) for both tree-level, $G_0^{R,A,K}$, and the corrected, $G^{K}_{1, 2-\text{type}}$, propagators in the limit (\ref{limit}, \ref{limit2}) (with the initial and corrected ON and AA, correspondingly). Then the leading corrections to the ON (the element of the corrected propagator) is contained in:
\begin{equation}\label{n2phi3}
    \begin{aligned}
        &n^{(2)}_{k}(T_0) \approx \frac{\lambda^2}{2}\int_{t_0}^{T_0}dy^{0} dz^{0}\sqrt{g(y^{0})g(z^{0})} \int\frac{d^{D-1}\boldsymbol{\vec{p}}}{(2\pi)^{D-1}}\frac{d^{D-1}\boldsymbol{\vec{s}}}{(2\pi)^{D-1}}\delta\left(\boldsymbol{\vec{k}}-\boldsymbol{\vec{p}}-\boldsymbol{\vec{s}}\right){f^{in}_{k}}^{*}(y^{0})f^{in}_{k}(z^{0})\times \\
        &\times\Bigg\{{f^{in}_{p}}^{*}(y^{0})f^{in}_{p}(z^{0}){f^{in}_{s}}^{*}(y^{0})f^{in}_{s}(z^{0})\times\Big[\left(1+n_{k}\right)\left(1+n_{p}\right)\left( 1+n_{s}\right)-n_{k}n_{p}n_{s}\Big]+ \\
        &+2{f^{in}_{p}}^{*}(y^{0})f^{in}_{p}(z^{0})f^{in}_{s}(y^{0}){f^{in}_{s}}^{*}(z^{0})\times\Big[\left(1+n_{k}\right)\left(1+n_{p}\right)n_{s}-n_{k}n_{p}\left(1+n_{s}\right)\Big]+ \\
        &+f^{in}_{p}(y^{0}){f^{in}_{p}}^{*}(z^{0})f^{in}_{s}(y^{0}){f^{in}_{s}}^{*}(z^{0})\times\Big[\left(1+n_{k}\right)n_{p}n_{s}-n_{k}\left(1+n_{p}\right)\left(1+n_{s}\right)\Big]\Bigg\}+ h.c.,
    \end{aligned}
\end{equation}
while for the AA is in:
\begin{equation}\label{k2phi3}
    \begin{aligned}
        &\kappa^{(2)}_{k}(T_0) \approx \frac{\lambda^2}{2}\int_{t_0}^{T_0}dy^{0} dz^{0}\sqrt{g(y^{0})g(z^{0})} \int\frac{d^{D-1}\boldsymbol{\vec{p}}}{(2\pi)^{D-1}}\frac{d^{D-1}\boldsymbol{\vec{s}}}{(2\pi)^{D-1}}\delta\left(\boldsymbol{\vec{k}}-\boldsymbol{\vec{p}}-\boldsymbol{\vec{s}}\right)f^{in}_{k}(y^{0})f^{in}_{k}(z^{0})\times \\
        &\times\Bigg\{f^{in}_{p}(y^{0}){f^{in}_{p}}^{*}(z^{0})f^{in}_{s}(y^{0}){f^{in}_{s}}^{*}(z^{0})\times \\        &\times\Big[\Big(\left(1+n_{p}\right)\left(1+n_{s}\right)+n_{p}n_{s}\Big)+sign\left(y^{0}-z^{0}\right)\left(1+2n_{k}\right)\Big(\left(1+n_{p}\right)\left(1+n_{s}\right)-n_{p}n_{s}\Big)\Big]+ \\
        &+2f^{in}_{p}(y^{0}){f^{in}_{p}}^{*}(z^{0}){f^{in}_{s}}^{*}(y^{0})f^{in}_{s}(z^{0})\times\\
        &\times\Big[\Big(\left(1+n_{p}\right)n_{s}+n_{p}\left(1+n_{s}\right)\Big)+sign(y^{0}-z^{0})\left(1+2n_{k}\right)\Big(\left(1+n_{p}\right)n_{s}-n_{p}\left(1+n_{s}\right)\Big)\Big]\Bigg\},
    \end{aligned}
\end{equation}
where $n_{k}$ is the initial value of the occupation number. Meanwhile initial $\kappa_k$ is assumed to be zero.

We are interested in the leading secular corrections from the last expressions to ON and AA, which means the largest correction as a function of $T_0=\frac{t_1+t_2}{2}$, when it is taken to the future infinity. To single out such contributions we perform the change of variables:
\begin{equation}\label{Tandtau}
    \tau=z_0-y_0,\ \ T=\frac{y_0+z_0}{2},
\end{equation}
Let us now divide the regions of the integration in (\ref{n2phi3}) and (\ref{k2phi3}) in to three intervals  $(t_{0}, \bar{t})$, $(\bar{t}, \widetilde{t})$, $(\widetilde{t}, T_0)$, where $\bar{t}\approx -\frac{1}{\rho} $ and $\widetilde{t}\approx \frac{1}{\rho}$. The first interval is the flat start of the expansion, where the modes behave as single oscillating exponents. The second interval is the transition between flat space and the eternal expansion. We can neglect contribution from this interval, because it provides subleading contribution in the limit in question: the duration of the interval is finite. The third interval is the expansion region, which becomes very large in the limit that we consider. On general grounds it can be expected that the largest contribution to AA and ON will come from the region of integration in (\ref{n2phi3}) and (\ref{k2phi3}) when both $z_0$ and $y_0$ lie in the third interval.

In fact, when either $y_0$ (or $z_0$) lies in the first interval while $z_0$ (or $y_0$) lies in the third interval we will obtain subleading oscillating correction to the propagator, because as $t\to - \infty$ the modes behave as single exponents with the frequency $k$ (\ref{modesas1}), while in the third region (as $t\to + \infty$) they behave as the superposition of exponents, oscillating for high momenta with frequency $\sqrt{k^2-\rho^2d^2}$ (or real for low momenta). Hence, in such a situation it is impossible to obtain non-oscillating contributions under the integrals on the RHS of (\ref{n2phi3}) and (\ref{k2phi3}). As a result there are no growing with time contributions, which come from these regions.

Another unimportant situation is when both $y_0$ and $z_0$ lie in the first interval. Here, the modes are almost free single plane waves. Also volume factor is approximately equal to unity. Hence, on the RHS of (\ref{n2phi3}) and (\ref{k2phi3}) we obtain the same expressions as in the flat space. The resulting largest contribution from grows linearly with time and has the form of the RHS of the Boltzmann's kinetic equation. Such a contribution is known to be equal to zero for the thermal (planckian) value ON. Moreover, such a secular growth has a simple physical meaning -- it describes the thermalization process: for any initial state (close to the equilibrium) we can take the initial Cauchy surface to the past infinity, $t_0 \rightarrow - \infty$, and assume that by the start of the expansion the AA are equal to zero and ON are equal to the thermal distribution \cite{Akhmedov:2021vfs}. 

Now, from (\ref{modesas1}) it can be seen that in the infinite future the behavior of the modes strongly differs for low and high momenta. Hence, we have to consider separately several situations depending on whether external or internal (loop) momenta are low or high.

\subsubsection{Diagrams of the second type with high external momenta}\label{2hem}

First, let us concentrate on the situation when the momentum of the external legs is large, i.e. obeys the condition $k > \rho d$. Naively it can be expected that leading loop corrections will come from the internal loop momenta also greater than $\rho d$, because processes of tunneling through the potential gap are suppressed. However, such naive arguments are not valid. For the integrity we will estimate and compare contributions from all values of momenta in the loop. We will show that small measure of integration in the interval  ($0$, $\rho d$) is not only compensated by the growth in the time of observation, but also contribution from this region of momentum space leads to much greater corrections to ON and AA. In the remainder of this subsection we will explicitly show the result only for ON, due to the similarity of the integrals in (\ref{n2phi3}) and (\ref{k2phi3}).

Consider the situation when both momenta in (\ref{n2phi3}) are high ($s,p>\rho d$). To obtain the leading contribution to (\ref{n2phi3}) in the limit (\ref{limit2}), we always can neglect every term from the product of the modes which gives oscillating functions of $T$  under the integral over $dT$ (see the definition of the notations in (\ref{Tandtau})). At the same time, after the integration over $\tau$ one will obtain a sum of $\delta$-functions in each contribution in (\ref{n2phi3}). Each $\delta$-function we can interpret as a kind of the energy conservation law\footnote{Note, however, that in such a background as (\ref{metric}) there is no exact energy conservation, because the modes are not simple exponents.}. Combing these ''energy conservation laws'' with the momentum conservation, one obtains three possible conditions for the presence of the secular contributions in (\ref{n2phi3}):
\begin{equation}\label{energycons2}
    \begin{aligned}
        &
        \begin{cases}
            \boldsymbol{\vec{k}} = \boldsymbol{\vec{p}}+\boldsymbol{\vec{s}} \\
            \sqrt{k^2-\rho^2d^2} = \sqrt{p^2-\rho^2d^2} +\sqrt{s^2-\rho^2d^2}
        \end{cases}, \\
        &
        \begin{cases}
            \boldsymbol{\vec{k}} = \boldsymbol{\vec{p}}+\boldsymbol{\vec{s}} \\
            \sqrt{k^2-\rho^2d^2} = \pm\sqrt{p^2-\rho^2d^2} \mp\sqrt{s^2-\rho^2d^2}
        \end{cases}.
    \end{aligned}
\end{equation}
Here, effectively masses are imaginary and the processes under considerations are possible, unlike the situation with the real mass. This is in accordance with the fact that there is no energy conservation in proper sense in time dependent backgrounds. After some simplifications and symmetrization one obtains that the leading correction to $n_{k}$ in the limit (\ref{limit}), (\ref{limit2}) for the large external momenta $k > \rho d$ is of the form:
\begin{equation}\label{n21phi3max}
\boxed{
    \begin{aligned}
        &n_{k \gg \rho d}^{(2, 1)}(T_0) \approx \frac{\pi\lambda^2}{16}e^{-\frac{D-6}{2}\rho T_0}\int_{p \geq \rho d}\frac{d^{D-1}\boldsymbol{\vec{p}}}{(2\pi)^{D-1}}\int_{p \geq \rho d}\frac{d^{D-1}\boldsymbol{\vec{s}}}{(2\pi)^{D-1}}\delta\left(\boldsymbol{\vec{k}}-\boldsymbol{\vec{p}}-\boldsymbol{\vec{s}}\right)\frac{1}{kps}\times\\
        &\times\Bigg\{\mathcal{N}_{1}(k, p, s)\times\Big[\left(1+n_{k}\right)\left(1+n_{p}\right)\left( 1+n_{s}\right)-n_{k}n_{p}n_{s}\Big]+ \\
        &+2\mathcal{N}_{2}(k, p, s)\times\Big[\left(1+n_{k}\right)\left(1+n_{p}\right)n_{s}-n_{k}n_{p}\left(1+n_{s}\right)\Big]+ \\
        &+\mathcal{N}_{3}(k, p, s)\times\Big[\left(1+n_{k}\right)n_{p}n_{s}-n_{k}\left(1+n_{p}\right)\left(1+n_{s}\right)\Big]\Bigg\}.
    \end{aligned}
}
\end{equation}
Here, we use the following notations:
\begin{equation}\label{Ns}
    \begin{aligned}
        &\mathcal{N}_{1}(k, p, s)=\delta\big(\omega(k)-\omega(p)-\omega(s)\big)\times\Big|D_{1}(k)D_{2}(p)D_{2}(s)+D_{2}(k)D_{1}(p)D_{1}(s)\Big|^2+ \\
        &+2\delta\big(\omega(k)+\omega(p)-\omega(s)\big)\times\Big|D_{1}(k)D_{1}(p)D_{2}(s)+D_{2}(k)D_{2}(p)D_{1}(s)\Big|^2,\\
        &\mathcal{N}_{2}(k, p, s)= \delta\big(\omega(k)-\omega(p)-\omega(s)\big)\times\Big|D_{1}(k)D_{2}(p)D_{1}^{*}(s)+D_{2}(k)D_{1}(p)D_{2}^{*}(s)\Big|^2+ \\
        &+\delta\big(\omega(k)+\omega(p)-\omega(s)\big)\times\Big|D_{1}(k)D_{1}(p)D_{1}^{*}(s)+D_{2}(k)D_{2}(p)D_{2}^{*}(s)\Big|^2+ \\
        &+\delta\big(\omega(k)-\omega(p)+\omega(s)\big)\times\Big|D_{1}(k)D_{2}(p)D_{2}^{*}(s)+D_{2}(k)D_{1}(p)D_{1}^{*}(s)\Big|^2,\\
        &\mathcal{N}_{3}(k, p, s)=\delta\big(\omega(k)-\omega(p)-\omega(s)\big)\times\Big|D_{1}(k)D_{1}^{*}(p)D_{1}^{*}(s)+D_{2}(k)D_{2}^{*}(p)D_{2}(s)\Big|^2+ \\
        &+2\delta\big(\omega(k)+\omega(p)-\omega(s)\big)\times\Big|D_{1}(k)D_{2}^{*}(p)D_{1}^{*}(s)+D_{2}(k)D_{1}^{*}(p)D_{2}^{*}(s)\Big|^2,
    \end{aligned}
\end{equation}
where coefficients $D_1$ and $D_2$ are defined in (\ref{modesfinal1}).

Before continuing our analysis, let us clarify a few things about the time dependence (on $T_0$) that we have obtained in (\ref{n21phi3max}). It is straightforward to see that the exponential growth of (\ref{n21phi3max}) comes only from two sources -- volume factor in every vertex and damping exponents from each mode. In such a situation it is easy to find the corresponding power of the exponent for arbitrary dimension and degree of interaction (when all momenta in the loop are higher than $\rho d$). The result for $\lambda\varphi^b$ theory in $D$ dimensions is the following:
\begin{equation}\label{n2phibmax0}
\boxed{
    \begin{aligned}
        &n^{(2)}_{k \gg \rho d}(T_0) \sim 
        \lambda^{2} e^{\rho\left(\frac{2D-b(D-2)}{2}\right)T_0}\Phi\left(\frac{k}{\Lambda}\right),
    \end{aligned}
}
\end{equation}
\begin{equation}\label{k2phibmax0}
\boxed{
    \begin{aligned}
        &\kappa^{(2)}_{k \gg \rho d}(T_0) \sim 
         \lambda^2 e^{\rho\left(\frac{2D-b(D-2)}{2}\right)T_0}\Phi\left(\frac{k}{\Lambda}\right),
    \end{aligned}
}
\end{equation}
where $b$ is the degree of the interaction. For example, $2D-b(D-2)=0$, when $D=4,\ b=4$. Then we get the linear growth in time, as the integrand of $dT$ in the generalization of (\ref{n2phi3}) and (\ref{k2phi3}) for $\lambda\varphi^b$ theory in $D$ dimensions is independent of $T$.  From this point of view it becomes clear that this factor in the exponent is connected to the renormabilty conditions of the $\lambda\varphi^b$ theory in $D$ dimensions. 

Now let us look at the contribution to (\ref{n2phi3}) and (\ref{k2phi3}) coming from the region when only one of the momenta in the loop is higher than $\rho d$, for example $p>\rho d$, and the other is smaller than $\rho d$. While the integration over small momenta in the loop integrals has small measure, the behaviour of the modes with small momenta radically differs from the modes with high momenta as can be seen from (\ref{modesas1}), (\ref{modesfinal1}). 

As the modes with momenta $s < \rho d$ do not oscillate,  from the integrals over $\tau$ one will approximately obtain terms with $\delta$-functions of the form $\delta(\omega_k-\omega_p)$ and $\delta(\omega_k+\omega_p)$, instead of terms with $delta$-functions with three frequencies as in (\ref{Ns}). Terms with $\delta(\omega_k+\omega_p)$ are always vanishing, while the other $\delta$-function imposes the equality $k=p$. Furthermore, one can neglect every term which contains exponent from the modes of momenta $s$ with negative power. Combining all these observations together one can write down the leading correction to ON from the integration region under consideration in the following form:

\begin{equation}\label{n2highextlowint}
    \begin{aligned}
        & n^{(2)}_{k > \rho d,\; p > \rho d,\; s\leq \rho d}(T_0) \approx \lambda^{2}e^{\rho T_0}\frac{2k}{\omega_{k}}\int \frac{d\Omega_{\boldsymbol{\vec{p}}}}{(2\pi)^3}\int_{0}^{s=\rho d}\frac{d^3\boldsymbol{\vec{s}}}{(2\pi)^3}\delta(\boldsymbol{\vec{k}}-\boldsymbol{\vec{p}}-\boldsymbol{\vec{s}})\frac{1}{s}\frac{|E_1(s)|^{2} e^{2 \sqrt{\rho^2d^2 - s^2} T_0}}{\rho+2\sqrt{\rho^2d^2 - s^2}} \times \\
        & \times\Bigg\{4|D_1(k)|^2|D_2(k)|^2(1+2n_k)(1+2n_s)-(|D_1(k)|^2+|D_2(k)|^2)^2n_{k}\left(1+n_{k}\right)\Bigg\},
    \end{aligned}
\end{equation}
where $d\Omega_{\boldsymbol{\vec{p}}}$ is the measure of integration over the angles of $\boldsymbol{\vec{p}}$. The obtained integral can be estimated by the steepest descent method, and one can see that such a contribution grows with time as $\frac{e^{2\rho T_0}}{\sqrt{T_0}}$. Such a time dependence is valid only for the in-vacuum state, namely when $n_k$ is equal to zero. 

However the situation becomes more complicated for the thermal initial distribution. As we work with the massless scalar field, the thermal distribution behaves as $n_{k}\approx \frac{1}{k}$ when $k\rightarrow 0$. Hence, the remaining integral in the expression under consideration, multiplying the factor $\frac{e^{2\rho T_0}}{\sqrt{T_0}}$, contains the IR divergence. The origin of this divergence is very similar to the one in the kinetic equation for the massless fields in flat space. The question of how to deal with such a divergence becomes critical, when one is going to perform the resummation of leading contributions from all loops. In such a situation the mode function renormalization from tadpole diagrams can play an important role. In any case this question demands a separate carefull study. 

The analysis for the situation when both $\boldsymbol{\vec{s}}$ and $\boldsymbol{\vec{p}}$ lie in the interval $(0,\rho d)$ is much more simple. Every term from the product ${f^{in}_{k}}^{*}(y^{0})f^{in}_{k}(z^{0})$  gives only an oscillating contribution. And these oscillations cannot be compensated by the modes of momenta $\boldsymbol{\vec{p}}$ or $\boldsymbol{\vec{s}}$, as the latter are real. Thus, the contribution from the region ($s,p\leq \rho d$) does not lead to a secular growth. 

In all, the contribution to ON with $k > \rho d$ from the region ($p,s>\rho d$) contains the growth of the form $e^{\rho T_0}$ (\ref{n21phi3max}), while the contribution from the region ($s\leq\rho d,\ p>\rho d$) contains the growth of the form $\frac{e^{2\rho T_0}}{\sqrt{T_0}}$ (\ref{n2highextlowint}). Thus, the largest contribution to the ON comes from the region where only one of the momenta in the loop is higher than $\rho d$. The situation with AA is similar. Such a result differs from the well studied cases, where the largest contribution for the high external momenta is coming from also high momenta in the loops \cite{Krotov:2010ma}, \cite{Akhmedov:2013vka}.

\subsubsection{Diagramms of the second type with low external  momenta }\label{2lem}

Let us concentrate now on the case when the momentum in the external leg in (\ref{n2phi3}) and (\ref{k2phi3}) obeys $k \leq \rho d$. First, let us consider contribution to ON from the high momenta in the loop, $s,p>\rho d$. To obtain the leading contribution to (\ref{n2phi3}) in the limit (\ref{limit2}) from such a region of integration we can neglect every term from the product of modes which contains oscillating functions of $T$ (see (\ref{Tandtau}) for the definition) under the $dT$ integral. Then, in the leading terms there will be present $\delta(\omega_s-\omega_p)$, appearing after the integration over $\tau$. In the resulting expression the time dependence is contained only in the volume factors and external leg modes. So the corresponding correction grows with time in the following way:

\begin{equation}
   n^{(2)}_{k \leq \rho d,\; p, s > \rho d}(T_0)\propto e^{\rho T_0}e^{2\sqrt{\rho^2d^2-k^2}T_0}.
\end{equation}
Second, consider the contribution from the region of integration when only one of the loop momenta, for example $\boldsymbol{\vec{p}}$, is greater than $\rho d$. But this situation is similar to the situation when $k>\rho d,\ s,p\leq \rho d$. There will be only oscillating contributions under the integral in (\ref{n2phi3}) and (\ref{k2phi3}). As a result, this region of integration does not bring any growing with $T_0$ corrections.

Finally, the largest contribution to (\ref{n2phi3}) and (\ref{k2phi3}) comes from the region of integration where all momenta, $k,p,s$, are lower than the $\rho d$ bound. For such a case we will also show the largest contribution to AA (apart from the one to ON) as this is the main result. In the situation under consideration the integration over $y_0$ and $z_0$ is especially simple, as there are no any oscillations, because all the modes in the integral are real exponential functions. As a result, after the integration over $dT$ one obtains that:
\begin{equation}
    \begin{aligned}
        & n^{(2)}_{k \leq \rho d,\; p,s\leq\rho d}(T_0)\approx \lambda^2e^{\rho T_0}e^{2\sqrt{\rho^2d^2-k^2}T_0}\frac{|E_{1}(k)|^{2}}{2k}\times\\
        & \times\int \frac{d^3\boldsymbol{\vec{p}}}{(2\pi)^3}\frac{d^3\boldsymbol{\vec{s}}}{(2\pi)^3}\delta(\boldsymbol{\vec{k}}-\boldsymbol{\vec{p}}-\boldsymbol{\vec{s}})e^{2\sqrt{\rho^2d^2-s^2}T_0}e^{2\sqrt{\rho^2d^2-p^2}T_0} H(k,p,s),
    \end{aligned}
\end{equation}
where $H(k,p,s)$ is a function depending only on $k,p,s$. We will specify this function in the final expression below. The only important fact at this point is that the function $e^{2\sqrt{\rho^2d^2-s^2}T_0}e^{2\sqrt{\rho^2d^2-p^2}T_0}$ is very fast in the limit (\ref{limit2}). So the corresponding integral can be estimated by the steepest descent. To do that let us make the change of variables:
\begin{equation}
    \boldsymbol{\vec{l}}=\frac{\boldsymbol{\vec{p}}-\boldsymbol{\vec{s}}}{2},\; \boldsymbol{\vec{L}}=\boldsymbol{\vec{p}}+\boldsymbol{\vec{s}}. 
\end{equation}
Then after the integration over $\boldsymbol{\vec{L}}$ we obtain that:
\begin{equation}\label{alllow}
    \begin{aligned}
        & n^{(2)}_{k \leq \rho d,\; p,s\leq\rho d}(T_0)\approx \lambda^{2}e^{\rho T_0}e^{2\sqrt{\rho^2d^2-k^2}T_0}\frac{|E_1(k)|^{2}}{2k}\times \\
        & \int\frac{d^3\boldsymbol{\vec{l}}}{(2\pi)^3}e^{2\sqrt{\rho^2d^2-\left(\frac{\boldsymbol{\vec{k}}}{2}+\boldsymbol{\vec{l}}\right)^2}T_0}e^{2\sqrt{\rho^2d^2-\left(\frac{\boldsymbol{\vec{k}}}{2}-\boldsymbol{\vec{l}}\right)^2}T_0} H(k,l)=\lambda^{2}\frac{e^{\rho T_0}e^{2\sqrt{\rho^2d^2-k^2}T_0}e^{4\sqrt{\rho^2d^2-\frac{k^2}{4}}T_0}}{T_0^{\frac{3}{2}}}\times \\
        & \times\frac{|E_1(k)|^{2}\left|E_1\left(\frac{k}{2}\right)\right|^4\left(1+2n_{\frac{k}{2}}\right)^2}{2k^3\left(\frac{\rho}{2}+\sqrt{\rho^2d^2-k^2}+2\sqrt{\rho^2d^2-\frac{k^2}{4}}\right)^2}\sqrt{\frac{\left(\rho^2d^2-\frac{k^2}{4}\right)^{\frac{9}{2}}}{8\left(\rho^2d^2+\frac{k^2}{2}\right)}}.
    \end{aligned}
\end{equation}
Similarly for AA one can obtain the expression as follows:
\begin{equation}\label{kappaalllow}
    \begin{aligned}
        & \kappa^{(2)}_{k \leq \rho d,\; p,s\leq\rho d}(T_0)=\lambda^{2}\frac{e^{\rho T_0}e^{2\sqrt{\rho^2d^2-k^2}T_0}e^{4\sqrt{\rho^2d^2-\frac{k^2}{4}}T_0}}{T_0^{\frac{3}{2}}}\times \\
        & \times\frac{E^{2}_{1}(k)\left|E_{1}\left(\frac{k}{2}\right)\right|^{4}\left[\left(1+2n_{\frac{k}{2}}\right)^2+2\left(1+n_{\frac{k}{2}}\right)n_{\frac{k}{2}}\right]}{4k^3\left(\frac{\rho}{2}+\sqrt{\rho^2d^2-k^2}+2\sqrt{\rho^2d^2-\frac{k^2}{4}}\right)^2}\sqrt{\frac{\left(\rho^2d^2-\frac{k^2}{4}\right)^{\frac{9}{2}}}{8\left(\rho^2d^2+\frac{k^2}{2}\right)}}.
    \end{aligned}
\end{equation}
It is not hard to see that these contributions are the largest in (\ref{n2phi3}) and (\ref{k2phi3}) in the limit that we consider.

Let us close this section by summing up our observations: we have calculated and analyzed time dependence of the corrections to the ON ($n_{k}(T_0)$ from (\ref{n2phi3})) and the AA ($\kappa_{k}(T_0)$ from (\ref{k2phi3})) in the $\frac{\lambda}{3!}\phi^{3}$-theory in $D$ dimensions. Here $T_0$ is the average time of the corrected Keldysh propagator. Corrections to ON and AA for large external momenta $k > \rho d$ coming from the large internal loop momenta $p,s > \rho d$ are of the form (\ref{n21phi3max}).  They are both exponentially suppressed for $D > 6$ and  grow for $D<6$. In any case this is a subleading correction. Namely, we have shown that  the largest contribution for high external momenta comes from the low internal momenta (\ref{n2highextlowint}), and grows as $\frac{e^{2\rho T_0}}{\sqrt{T_0}}$. The same analysis was made for the corrections to the $n_{k}(T_0)$ and $\kappa_{k}(T_0)$ for low external momenta $k\leq \rho d$. We have shown that the largest contribution in this case comes from the low internal momenta $p,s\leq\rho d$ and is given by the expressions (\ref{alllow}) and (\ref{kappaalllow}). In other words, the largest growing with $T_0$ contribution to ON and AA goes to $k < \rho d$ and it comes also from low momenta in the loop $p,s < \rho d$. This observation will drastically simplify the resummation of the leading corrections from all loops.  


\end{section}

\begin{section}{The expectation value of the stress-energy tensor}\label{sec5}

In this section we analyse how loop corrections to ON and AA affect the expectation value of the stress-energy tensor (SET). We compare tree-level value of this quantity with loop-corrections to it. For simplicity in this section the initial state is taken to be the in-vacuum. The SET operator is defined as follows: 
\begin{equation}
    \begin{aligned}
        & T_{\mu\nu}(t, \boldsymbol{\vec{x}}) \equiv \frac{2\delta\mathscr{L}[\phi, g]}{\sqrt{|g|}\delta g^{\mu\nu}} = \\
        & = \partial_{\mu}\phi(t, \boldsymbol{\vec{x}})\partial_{\nu}\phi(t, \boldsymbol{\vec{x}}) - \frac{g_{\mu\nu}}{2}\Big[g^{\alpha\beta}\partial_{\alpha}\phi(t, \boldsymbol{\vec{x}})\partial_{\beta}\phi(t, \boldsymbol{\vec{x}})\Big].
    \end{aligned}
\end{equation}
The interaction term provides a subleading contribution due to the presence of $\lambda$, which is assumed to be small. Then the expectation value of the SET can be found as follows:
\begin{equation}\label{SETfromG}
    \expval{T_{\mu\nu}(t, \boldsymbol{\vec{x}})}=\lim_{\boldsymbol{\vec{x}_{1}}\rightarrow \boldsymbol{\vec{x}_{2}}}\left[\frac{\partial}{\partial x^{\mu}_1}\frac{\partial}{\partial x^{\nu}_2}-\frac{\eta_{\mu\nu}}{2}\eta^{\alpha\beta}\frac{\partial}{\partial x^{\alpha}_1}\frac{\partial}{\partial x^{\beta}_2}\right] G^{K}\left(t, \boldsymbol{\vec{x}_{1}}, t, \boldsymbol{\vec{x}_{2}}\right).
\end{equation}
We use space-like point splitting to obtain the real result.

Loop corrected ON and AA are changing in time as fast as the mode functions. Hence, we can not neglect time derivatives of $n_{k}$ and $\kappa_{k}$ in the expression for the SET. This is a quite unusual situation in comparison with the standard one in kinetic processes. 

To calculate the SET let us rewrite the expression for the tree-level Keldysh propagator in the following form:
\begin{equation}\label{treedamp}
    \begin{aligned}
        & G^{K}_{0}(t, \boldsymbol{\vec{x}_{1}}, t, \boldsymbol{\vec{x}_{2}})=\int \frac{d^3\boldsymbol{\vec{k}}}{(2\pi)^3}\left[{f_{k}^{in}}^{*}(t)f_{k}^{in}(t)\right]e^{i\boldsymbol{\vec{k}}(\boldsymbol{\vec{x}_{1}}-\boldsymbol{\vec{x}_{2}})}\approx \\ 
        & \approx \int_{k<\rho d} \frac{d^3\boldsymbol{\vec{k}}}{(2\pi)^3}\left[A_1(k)e^{-\rho  t}e^{2 \sqrt{\rho^2d^2 - k^2} t}+A_2(k)e^{-\rho  t}+A_3(k)e^{-\rho  t}e^{-2 \sqrt{\rho^2d^2 - k^2} t}\right]e^{i\boldsymbol{\vec{k}}(\boldsymbol{\vec{x}_{1}}-\boldsymbol{\vec{x}_{2}})}+\\
        & +\int_{k>\rho d}\frac{d^3\boldsymbol{\vec{k}}}{(2\pi)^3}\left[B_1(k)e^{-\rho  t}e^{2i \sqrt{ k^2-\rho^2d^2} t}+B_2(k)e^{-\rho  t}+B_3(k)e^{-\rho  t}e^{-2i \sqrt{k^2-\rho^2d^2} t}\right]e^{i\boldsymbol{\vec{k}}(\boldsymbol{\vec{x}_{1}}-\boldsymbol{\vec{x}_{2}})},
    \end{aligned}
\end{equation}
where the coefficients $A_n(k)$ and $B_n(k)$, $n=1,2,3$ are independent of $t$ and $\boldsymbol{\vec{x}}$ and have the following form:
\begin{equation}
    \begin{aligned}
        & A_1(k)=\frac{1}{2k} |E_1(k)|^2, \ A_2(k)=\frac{1}{2k}\left[E_1(k)E^{*}_2(k)+E_1^{*}(k)E_2(k)\right], \ A_3(k)=\frac{1}{2k} |E_2(k)|^2,\\
        & B_1(k)=\frac{1}{2k} D_1(k)D_2^{*}(k), \ B_2(k)=\frac{1}{2k}\left[|D_1(k)|^2+|D_2(k)|^2\right], \ B_3(k)=\frac{1}{2k} D_1^{*}(k)D_2(k),
    \end{aligned}
\end{equation}
Where $E_{1,2}(k)$ and $D_{1,2}(k)$ are defined in (\ref{modesfinal1}).
 
As exponents are eigen-functions of the derivative operator, the time dependence of the Fourier image of the SET is essentially the \textcolor{black}{same} as of the propagator (\ref{treedamp}). Hence, we can straightforwardly see that the tree-level expectation value of the SET is decaying to zero in the future. This can be expected on general grounds due to the expansion of the space-time background.

However, in the sections \ref{2hem}, \ref{2lem} we have shown that loop corrections to ON and AA grow with time exponentially. These quantities are elements of the Keldysh propagator in (\ref{SETfromG}) and they grow with the average time of this propagator. Furthermore, the time dependence of each term in the Fourier image of the SET coincides with time dependence of the corresponding term from the Fourier image of the Keldysh propagator $G_K$. We have found the largest correction to ON and AA and, hence, to $G_K$ in the previous section. Then we can single out the leading loop contribution to the SET (it comes from the low momenta modes):

\begin{equation}\label{loopgrow}
    \expval{T_{\mu\nu}(t)}_{1-\text{loop}}\approx\int_{k<\rho d} \frac{d^3\boldsymbol{\vec{k}}}{(2\pi)^3}C_{1}(k)\frac{e^{4\sqrt{\rho^2d^2-k^2}t}e^{4\sqrt{\rho^2d^2-\frac{k^2}{4}}t}}{t^{\frac{3}{2}}}K_{\mu\nu},
\end{equation}
where we have neglected derivatives of the function $\frac{1}{t^{\frac{3}{2}}}$, while the tensor $K_{\mu\nu}$ has the following form:
\begin{equation}
    \begin{aligned}
        & K_{\mu\nu} = \\
        & =\text{diag}\Bigg[8\left(\sqrt{\rho^2d^2-k^2}+\sqrt{\rho^2d^2-\frac{k^2}{4}}\right)^2+\frac{k^2}{2}, k^2_x-8\left(\sqrt{\rho^2d^2-k^2}+\sqrt{\rho^2d^2-\frac{k^2}{4}}\right)^2+\frac{k^2}{2}, \\
        & k^2_y-8\left(\sqrt{\rho^2d^2-k^2}+\sqrt{\rho^2d^2-\frac{k^2}{4}}\right)^2+\frac{k^2}{2}, k^2_z-8\left(\sqrt{\rho^2d^2-k^2}+\sqrt{\rho^2d^2-\frac{k^2}{4}}\right)^2+\frac{k^2}{2}\Bigg],
    \end{aligned}
\end{equation}
and $C_{1}$ is :
\begin{equation}
    \begin{aligned}
        & C_{1}(k)=\lambda^{2}\frac{|E_{1}(k)|^{4}\left|E_{1}\left(\frac{k}{2}\right)\right|^4\left(1+2n_{\frac{k}{2}}\right)^2}{2k^3\left(\frac{\rho}{2}+\sqrt{\rho^2d^2-k^2}+2\sqrt{\rho^2d^2-\frac{k^2}{4}}\right)^2}\sqrt{\frac{(\rho^2d^2-\frac{k^2}{4})^{\frac{9}{2}}}{8(\rho^2d^2+\frac{k^2}{2})}}+\\
        & +\lambda^{2}
       \frac{\left|E_{1}\left(\frac{k}{2}\right)\right|^4\left[\left(1+2n_{\frac{k}{2}}\right)^2+2\left(1+n_{\frac{k}{2}}\right)n_{\frac{k}{2}}\right]}{4k^3(\frac{\rho}{2}+\sqrt{\rho^2d^2-k^2}+2\sqrt{\rho^2d^2-\frac{k^2}{4}})^2}\sqrt{\frac{(\rho^2d^2-\frac{k^2}{4})^{\frac{9}{2}}}{8(\rho^2d^2+\frac{k^2}{2})}}\left[E^{4}_{1}(k) + {E^{4}_{1}(k)}^{*}\right].    
    \end{aligned}
\end{equation}
Now we can compare the time dependence of the tree-level expectation value of the SET with the loop corrected expectation value. The loop corrected expectation value (\ref{loopgrow}) grows with time, while the tree-level expectation value is decaying towards future infinity. Hence, loop corrections in such a situation are extremely significant: they do not just go into coupling constants renormalizations, but also modify the energy fluxes. To get the complete understanding of the physics in such a situation one has to resum the largest growing contributions from all perturbative orders. Obtained result, as was mentioned, has a different character from the kinetic one. Hence, resummation of all perturbative orders in this situation is more complicated than in the standard case. Problem of this kind we will try to solve in the following papers. 

\end{section}

\begin{section}{Conclusion and acknowledgements}\label{sec6}

We have considered quantum loop corrections to the occupation number and anomalous quantum average in $\lambda\varphi^3$ theory in $D$ dimensions. (We assume that stabilizing $\phi^4$ terms are also present in the potential, but with much smaller coupling constants and correct our observations at much larger time scales.) The theory was considered in a FLRW space-time with flat start at past infinity and the initial state was taken as the in- Fock space ground state. 

We have show that the contribution of \textcolor{black}{tadpole} diagrams can be absorbed into a change of the mode functions and mass renormalization. At the same time diagrams of the second type (shown on the Fig. \ref{2type}) lead to a change of the state of the theory -- to a change in time of the occupation number and of the anomalous quantum average according to (\ref{n2phi3}) and (\ref{k2phi3}). We look for the largest contributions to these expressions in the limit (\ref{limit}), (\ref{limit2}), when both points of the propagator are taken to the future infinity. The growth of these quantities with the average time $T_0$ of the Keldysh propagator comes from the region of the expansion of the space-time, if the initial state is stationary: the flat start and transition regions in the FLRW space-time contribute subleading corrections. The change of the occupation number and of the anomalous quantum average in time shows the change of the state of the theory during the course of its evolution.

We have shown that the largest contribution to the occupation number and anomalous quantum average with (external) momenta $k>\rho d$ comes from the region where only one of the (internal) momenta in the loops is higher than the $\rho d$ bound. Such a result is quite counter-intuitive: on general grounds it can be expected that the largest contribution for high external momenta should come also from high momenta in the loops. At least that is the case in de Sitter space-time \cite{Krotov:2010ma}, \cite{Akhmedov:2013vka}. Such an unusual phenomenon we attribute to the specific behavior of the modes for low momenta. 

Finally, we have shown that the fastest possible growth appears in occupation number and anomalous quantum average for low  (external) momenta, $k \leq\rho d$, and comes also from the low (internal) momenta in the loops, $p,s\leq\rho d$. In this case for future references we find the expressions with all explicit coefficients in (\ref{alllow}) and (\ref{kappaalllow}). The dependence on time is not of a kinetic type, which complicates the situation with the resummation of loops for generic initial conditions.

We show that the change of the state of the theory that we observe cannot be neglected as the loop corrected stress-energy tensor (\ref{loopgrow}) is much larger the tree-level one in the future infinity. This result signals that for complete analysis of the theory in such a background we need to perform a resummation of the leading growing corrections from all perturbative orders. Our observations show that one can take care of only about the modes with low momenta. To perform the resummation we need to check the growth of the multiple point correlation functions and solve the (system) of Dyson-Schwinger equations.

We would like to acknowledge discussions with A.Alexandrov, K.Bazarov, D. Diakonov, K.Gubarev, A.Radkevich and A.Semenov. This work was supported by Russian Science Foundation (Project Number: 23-22-00145).
\end{section}

\addcontentsline{toc}{section}{References}\label{sec10}


\begin{thebibliography}{99}

\bibitem{Starobinsky:1980te}
A.~A.~Starobinsky,
Phys. Lett. B \textbf{91}, 99-102 (1980)
\href{https://doi.org/10.1016/0370-2693(80)90670-X}{doi:10.1016/0370-2693(80)90670-X}

\bibitem{Starobinsky:1982ee}
A.~A.~Starobinsky,
Phys. Lett. B \textbf{117}, 175-178 (1982)
\href{https://doi.org/10.1016/0370-2693(82)90541-X}{doi:10.1016/0370-2693(82)90541-X}

\bibitem{Linde:1981mu}
A.~D.~Linde,
Phys. Lett. B \textbf{108}, 389-393 (1982)
\href{https://doi.org/10.1016/0370-2693(82)91219-9}{doi:10.1016/0370-2693(82)91219-9}

\bibitem{Linde:1983gd}
A.~D.~Linde,
Phys. Lett. B \textbf{129}, 177-181 (1983)
\href{https://doi.org/10.1016/0370-2693(83)90837-7}{doi:10.1016/0370-2693(83)90837-7}

\bibitem{Guth:1980zm}
A.~H.~Guth,
Phys. Rev. D \textbf{23}, 347-356 (1981)
\href{https://doi.org/10.1103/PhysRevD.23.347}{doi:10.1103/PhysRevD.23.347}

\bibitem{Guth:1982ec}
A.~H.~Guth and S.~Y.~Pi,
Phys. Rev. Lett. \textbf{49}, 1110-1113 (1982)
\href{https://doi.org/10.1103/PhysRevLett.49.1110}{doi:10.1103/PhysRevLett.49.1110}

\bibitem{Albrecht:1982wi}
A.~Albrecht and P.~J.~Steinhardt,
Phys. Rev. Lett. \textbf{48}, 1220-1223 (1982)
\href{https://doi.org/10.1103/PhysRevLett.48.1220}{doi:10.1103/PhysRevLett.48.1220}

\bibitem{Akhmedov:2021rhq}
E.~T.~Akhmedov,
Mod. Phys. Lett. A \textbf{36}, no.20, 2130020 (2021)
\href{https://doi.org/10.1142/S0217732321300202}{doi:10.1142/S0217732321300202}
[\href{https://arxiv.org/abs/2105.05039}{arXiv:2105.05039} [gr-qc]].

\bibitem{Tsamis:1996qm}
N.~C.~Tsamis and R.~P.~Woodard,
Annals Phys. \textbf{253}, 1-54 (1997)
\href{https://doi.org/10.1006/aphy.1997.5613}{doi:10.1006/aphy.1997.5613}
[\href{https://arxiv.org/abs/hep-ph/9602316}{arXiv:hep-ph/9602316} [hep-ph]].

\bibitem{Tsamis:1996qq}
N.~C.~Tsamis and R.~P.~Woodard,
Nucl. Phys. B \textbf{474}, 235-248 (1996)
\href{https://doi.org/10.1016/0550-3213(96)00246-5}{doi:10.1016/0550-3213(96)00246-5}
[\href{https://arxiv.org/abs/hep-ph/9602315}{arXiv:hep-ph/9602315} [hep-ph]].

\bibitem{Tsamis:2005hd}
N.~C.~Tsamis and R.~P.~Woodard,
Nucl. Phys. B \textbf{724}, 295-328 (2005)
\href{https://doi.org/10.1016/j.nuclphysb.2005.06.031}{doi:10.1016/j.nuclphysb.2005.06.031}
[\href{https://arxiv.org/abs/gr-qc/0505115}{arXiv:gr-qc/0505115} [gr-qc]].

\bibitem{Krotov:2010ma}
D.~Krotov and A.~M.~Polyakov,
Nucl. Phys. B \textbf{849}, 410-432 (2011)
\href{https://doi.org/10.1016/j.nuclphysb.2011.03.025}{doi:10.1016/j.nuclphysb.2011.03.025}
[\href{https://arxiv.org/abs/1012.2107}{arXiv:1012.2107} [hep-th]].

\bibitem{Akhmedov:2011pj}
E.~T.~Akhmedov,
JHEP \textbf{01}, 066 (2012)
\href{https://doi.org/10.1007/JHEP01(2012)066}{doi:10.1007/JHEP01(2012)066}
[\href{https://arxiv.org/abs/1110.2257}{arXiv:1110.2257} [hep-th]].

\bibitem{Akhmedov:2013vka}
E.~T.~Akhmedov,
Int. J. Mod. Phys. D \textbf{23} (2014), 1430001
\href{https://doi.org/10.1142/S0218271814300018}{doi:10.1142/S0218271814300018}
[\href{https://arxiv.org/abs/1309.2557}{arXiv:1309.2557} [hep-th]].

\bibitem{Akhmedov:2015xwa}
E.~T.~Akhmedov, H.~Godazgar and F.~K.~Popov,
Phys. Rev. D \textbf{93} (2016) no.2, 024029
\href{https://doi.org/10.1103/PhysRevD.93.024029}{doi:10.1103/PhysRevD.93.024029}
[\href{https://arxiv.org/abs/1508.07500}{arXiv:1508.07500} [hep-th]].

\bibitem{Akhmedov:2019cfd}
E.~T.~Akhmedov, U.~Moschella and F.~K.~Popov,
Phys. Rev. D \textbf{99}, no.8, 086009 (2019)
\href{https://doi.org/10.1103/PhysRevD.99.086009}{doi:10.1103/PhysRevD.99.086009}
[\href{https://arxiv.org/abs/1901.07293}{arXiv:1901.07293} [hep-th]].

\bibitem{Akhmedov:2021agm}
E.~T.~Akhmedov, K.~V.~Bazarov and D.~V.~Diakonov,
Phys. Rev. D \textbf{104} (2021) no.8, 085008
\href{https://doi.org/10.1103/PhysRevD.104.085008}{doi:10.1103/PhysRevD.104.085008}
[\href{https://arxiv.org/abs/2106.01791}{arXiv:2106.01791} [hep-th]].

\bibitem{Akhmedov:2017dih}
E.~T.~Akhmedov and F.~Bascone,
Phys. Rev. D \textbf{97} (2018) no.4, 045013
\href{https://doi.org/10.1103/PhysRevD.97.045013}{doi:10.1103/PhysRevD.97.045013}
[\href{https://arxiv.org/abs/1710.06118}{arXiv:1710.06118} [hep-th]].

\bibitem{Trunin:2021lwg}
D.~A.~Trunin,
Phys. Rev. D \textbf{104} (2021) no.4, 045001
\href{https://doi.org/10.1103/PhysRevD.104.045001}{doi:10.1103/PhysRevD.104.045001}
[\href{https://arxiv.org/abs/2105.01647}{arXiv:2105.01647} [hep-th]].

\bibitem{Akhmedov:2023zfy}
E.~T.~Akhmedov, P.~S.~Zavgorodny, D.~I.~Sadekov and K.~A.~Kazarnovskii,
Phys. Rev. D \textbf{107}, no.12, 125006 (2023)
\href{https://doi.org/10.1103/PhysRevD.107.125006}{doi:10.1103/PhysRevD.107.125006}
[\href{https://arxiv.org/abs/2303.08624}{arXiv:2303.08624} [hep-th]].

\bibitem{Akhmedov:2022whm}
E.~T.~Akhmedov and P.~A.~Anempodistov,
Phys. Rev. D \textbf{105}, no.10, 105019 (2022)
\href{https://doi.org/10.1103/PhysRevD.105.105019}{doi:10.1103/PhysRevD.105.105019}
[\href{https://arxiv.org/abs/2204.01388}{arXiv:2204.01388} [hep-th]].

\bibitem{Birrell:1982ix}
N.~D.~Birrell and P.~C.~W.~Davies,
``Quantum Fields in Curved Space,''
\href{https://doi.org/10.1017/CBO9780511622632}{doi:10.1017/CBO9780511622632}









\bibitem{Kamenev}
A. Kamenev,
“Many-body theory of non-equilibrium systems,” Cambridge, UK: Univ. Pr. (2011)
[\href{https://arxiv.org/abs/cond-mat/0412296}{arXiv:cond-mat/041229}].

\bibitem{Berges:2004yj}
J.~Berges,
AIP Conf. Proc. \textbf{739}, no.1, 3-62 (2004)
\href{https://doi.org/10.1063/1.1843591}{doi:10.1063/1.1843591}
[\href{https://arxiv.org/abs/hep-ph/0409233}{arXiv:hep-ph/0409233} [hep-ph]].






\bibitem{Akhmedov:2021vfs}
E.~T.~Akhmedov and K.~Kazarnovskii,
Universe \textbf{8}, no.3, 162 (2022)
\href{https://doi.org/10.3390/universe8030162}{doi:10.3390/universe8030162}
[\href{https://arxiv.org/abs/2110.00454}{arXiv:2110.00454} [hep-th]].




\end{thebibliography}
\end{document}